\documentclass[aps,prb]{revtex4}
\usepackage{graphicx}
\usepackage{bm}
\usepackage{xfrac}
\usepackage{amsmath}
\usepackage{amssymb}
\usepackage{ytableau}
\usepackage{multirow}

\newcommand{\be}{\begin{equation}}
\newcommand{\ee}{\end{equation}}
\newcommand{\bea}{\begin{eqnarray}}
\newcommand{\eea}{\end{eqnarray}}
\newcommand{\beb}{\begin{eqnarray*}}
\newcommand{\eeb}{\end{eqnarray*}}

\newcommand{\phrb}[3]{Phys. Rev. B{\bf #1}, #2 (#3).}

\begin{document}
\title{Quantum Hall Skyrmions at $\nu=0,\pm 1$ in monolayer graphene}

\author{Thierry Jolicoeur$^{1}$}
\author{Bradraj Pandey$^2$}

\affiliation{$^1$Institut de Physique Th\'eorique, CNRS, Universit\'e Paris-Saclay,
91190 Gif sur Yvette, France}

\affiliation{$^2$Laboratoire de Physique Th\'eorique et Mod\`eles statistiques,
CNRS,
Universit\'e Paris-Saclay, 91405 Orsay, France}

\date{2019}
\begin{abstract}
Monolayer graphene under a strong perpendicular magnetic field exhibit quantum Hall ferromagnetism
with spontaneously broken spin and valley symmetry~\cite{nomura}. The approximate SU(4) spin/valley symmetry is
broken by small lattice scale effects in the central Landau level corresponding to filling factors
$\nu=0,\pm 1$. Notably the ground state at $\nu=0$ is believed to be a canted antiferromagnetic (AF) or a ferromagnetic (F)
state depending on the component of the magnetic field parallel to the layer and the strength of small
 anisotropies. We study the skyrmions for the filling factors $\nu=\pm 1,0$ by using
exact diagonalizations on the spherical geometry.
If we neglect anisotropies we confirm the validity of the standard skyrmion picture generalized to four degrees of freedom.
For filling factor $\nu=- 1$ the hole skyrmion is an infinite-size valley skyrmion with full spin polarization
because it does not feel the anisotropies. The electron skyrmion is also always of infinite size. In the F phase
it is always fully polarized while in the AF phase it undergoes continuous magnetization under increasing Zeeman energy.
In the case of $\nu=0$ the skyrmion is always maximally localized in space both in F and AF phase.
In the F phase it is fully polarized while in the AF phase it has also progressive magnetization with Zeeman energy.
The magnetization process is unrelated to the spatial profile of the skyrmions contrary to the SU(2) case.
In all cases the skyrmion physics is dominated by the competition between anisotropies and Zeeman effect
but not directly by the Coulomb interactions, breaking universal scaling with the ratio Zeeman to Coulomb energy.

\end{abstract}
\maketitle
\section{introduction}
Monolayer graphene is a new two-dimensional (2D) electron system with 
Dirac cones in the energy-band structure. 
Application of a perpendicular magnetic field leads to the formation of Landau levels that have peculiar features
not seen in conventional 2D electron gases in semiconducting materials like GaAs. Notably~\cite{graphene_exp,novoselov,zhang,zhang2,Barlas12}
at neutrality there is a set of fourfold degenerate Landau levels that stay at zero energy for all
values of the field in the absence of Zeeman energy. The graphene physics in this regime is dominated by the interplay
between the twofold valley degeneracy and the usual spin degeneracy. This special feature
has been clearly observed experimentally~\cite{graphene_exp,novoselov,zhang,zhang2}. 
It has been shown notably that the fourfold degeneracy is completely lifted leading to quantum Hall states
at Landau level filling factors $\nu=0,\pm 1$.
This phenomenon can be described as a
generalization of the quantum Hall ferromagnetism~\cite{Yang2006} which is well known in the case
of the spin degree of freedom~\cite{rezayi87,rezayi91,Sondhi93,Moon95,Yang96,Abolfath97,fertig,Fertig97,Cooper97,skyrnmr,sondhiwu} 
or in the ``which layer'' pseudospin for bilayer
2D electron gases~\cite{gm,sds,skyrexpt}. In the case of graphene there is an approximate SU(4) symmetry which is spontaneously broken
giving rise to several possible ground states and associated collective Goldstone modes~\cite{Yang96}.

A general consequence of the formation of quantum Hall ferromagnetic ground states is that charge carriers
are now skyrmions instead of simple quasiparticles describable by Hartree-Fock type electrons or holes.
The local extra charge positive or negative is accompanied by a texture in the internal degrees of
freedom i.e. spin and valley in the case of graphene. These entities are well-known to appear in the case of SU(2) degeneracy
but the graphene system gives them an even richer 
structure~\cite{Arovas99,Doretto2007,Shibata2008,Abanin2013,ezawa03,ezawa05,tsit,Toke2012,Lian1,Lian2}. 
This is expected to happen close to all 
integer filling factors. In the case of spin skyrmions there is a competition between Coulomb interactions 
and the Zeeman energy~: indeed the interactions favor an extended texture while the Zeeman energy favor
localized states. As a consequence there is a finite number of spin flips in a given texture which can be
tuned by varying the ratio of these two energies. This interplay can be captured by 
Hartree-Fock theory\cite{Brey96,BreyL,YS,AGG,ERS,SGJ,Sanka,Ferrer,Bar,mfb,elandholeenergy}
and exact diagonalization techniques~\cite{Xie96,Palacios96,AHM98,Wojs2002}.
In the case of valley skyrmions there is no analog
of the Zeeman energy so exchange interactions favor infinitely large skyrmions~\cite{mansour}. The case of graphene
is richer due to the fact that one may have both spin flips and valley flips. Previous works have investigated
this competition for various filling factors. However there is a crucial ingredient that has to be included~:
the fact that the SU(4) symmetry is explicitly broken by lattice-scale effects~\cite{alicea,GMD,Khari1,Khari2,Wu}. 
This phenomenon is particularly acute
in the central set of states $\nu=\pm 1,0$. Indeed at $\nu =0$ there is a huge degeneracy of SU(4) ground states.
To construct a Slater determinant describing $\nu=0$ according to quantum Hall ferromagnetism theory one has 
to select two orthogonal spinors in a 4-dimensional space spanned by spin and valley degrees of freedom. 
This leads~\cite{Igor,Jeil,Khari1,Khari2,Wu} to candidate states with ferromagnetic (F) or
antiferromagnetic (AF) ordering, charge-density wave or K\'ekul\'e states all related by the SU(4) symmetry.
It is likely that the ground state of monolayer graphene has antiferromagnetic order and that
the application of a field parallel to the plane by e.g. tilting the sample leads to a transition
to the ferromagnetic phase and a change of edge conduction~\cite{Young2012,Maher2013,Khari1,Khari2,Young2014}. 
It is thus important to understand the nature of charge carriers
at least in these two phases of graphene in the quantum Hall regime.

In this paper we use exact diagonalization of small systems in the spherical geometry to study the skyrmion physics.
Indeed it is known that the formation of skyrmions in usual SU(2) systems has very simple observable consequences
in this geometry. We show that in the SU(4) case there are no low-energy states beyond those associated with
skyrmions generalized to 4 flavor case. This is established by formulating counting rules for quantum skyrmion states.
We next add anisotropic terms breaking the valley symmetry that are induced by lattice-scale physics.
They can be written in a simple form which is parameterized by only two unknown quantities~: in addition to an overall
strength of the anisotropies we use an angle whose variation allows to study both AF and F phases.

In the case of filling factor $\nu=-1$ we investigate in detail the electron skyrmion since its hole counterpart 
does not feel the anisotropies.
We find that it is always of infinite size since the ground state has always zero total angular momentum and hence 
is delocalized over the whole sphere.
In the F phase it is also fully spin polarized but valley totally unpolarized. On the contrary in the AF phase
it undergoes progressive magnetization when increasing the Zeeman energy.
In the case of $\nu=0$ the skyrmion ground state in both phases F and AF prefers to have the maximum angular momentum
along the skyrmion branch of states meaning that it is of smallest possible spatial extent.
Again in the F phase it is fully polarized in spin and valley unpolarized while the AF phase also has progressive
magnetization and also no valley polarization.

In both cases the specific feature of these SU(4) skyrmions is that the magnetization process is determined
by the energy splitting entirely due to anisotropies within a single orbital multiplet~: $L=0$ in the case
of $\nu=\pm 1$ or $L=L_{max}$ for $\nu=0$. This splitting is ruled by the anisotropy energy scale and not by the Coulomb energy scale.
Hence it is a competition between Zeeman effect and anisotropies which governs skyrmion physics.

In section II we recall some relevant facts about monolayer graphene under a magnetic field and the anisotropies
important to the physics of the central zero-energy Landau level. In section III we discuss the physics of SU(2)
skyrmions in the sphere geometry.
In section III we concentrate on the filling factors $\nu=\pm 1$. Section IV is devoted to the filling $\nu=0$.
Finally section V contains our conclusions.

\section{QHE in monolayer graphene and symmetries}
When monolayer graphene is subjected to a perpendicular magnetic field $B$ there is formation of Landau levels
which have a simple form close to neutrality~:
\be
E_{ns}=sgn(n) \frac{\hbar v_F\sqrt{2}}{\ell_B}\sqrt{|n|} -\frac{1}{2}g_L \mu_B Bs,
\label{LLs}
\ee
where $n$ is an integer (positive or negative), the electronic spin is $s=\pm 1$,
the magnetic length is given by $\ell_B=\sqrt{\hbar c/eB}$, $v_F$ is the velocity at the Dirac point,
 $\mu_B$ the Bohr magneton, and the Land\'e factor $g_L=2$. There is an additional twofold degeneracy due
 to the two valleys $K$ and $K^\prime$. Electric neutrality of the graphene layer corresponds to half-filling of
 the zero-energy Landau level with $n=0$ in Eq.(\ref{LLs}). If we neglect the Zeeman energy $g_L \mu_B B$ 
 there is a fourfold degeneracy. Simple band filling of these Landau levels
 leads to the prediction of integer quantum Hall effect (IQHE) with Hall conductance~:
 \be
 \sigma_{xy}=(4p+2)\frac{e^2}{h},
 \ee
with $p$ positive or negative so IQHE occurs at filling factors $\nu=\pm 2, \pm 6, \pm 10,\dots$. Experiments 
on high-quality samples have revealed that in fact IQHE occurs at \textit{all} integer filling factors including
the special $\nu=0$ state. The states at fillings $\nu=\pm 1,0$ all correspond to some occurrence of quantum Hall 
ferromagnetism and are the subject of our study. We thus concentrate on the Fock space spanned by the orbital
$n=0$ set of states in the limit of negligible Landau level mixing.
Coulomb interactions between electrons is the essential ingredient for quantum Hall ferromagnetism. It is invariant
under SU(2) spin rotations but in fact there is a larger SU(4) invariance under unitary transformations in
the four-dimensional space spanned by the spin and valley degrees of freedom.
So the Hamiltonian we consider contains at least Coulomb interactions and Zeeman energy~:
\be
\mathcal{H}_0=\sum_{i<j}\frac{e^2}{\epsilon |\mathbf{r}_i-\mathbf{r}_j|} +\epsilon_Z\sum_i \sigma^z_i ,
\label{H0}
\ee
where in the Zeeman term we have used the Pauli matrix $\sigma_z$ in spin space and $\epsilon_Z=g_L\mu_B B/2$.
This symmetry however is only approximate and notably the fate of the ground states at $\nu=0$ depends
on details of the graphene system beyond this simple symmetric treatment. The short-distance behavior of
Coulomb interactions as well as electron-phonon interactions breaks down SU(4) in a way that can plausibly be modeled
by a simple local effective Hamiltonian as proposed notably by Kharitonov~\cite{Khari1,Khari2}:
\be
\mathcal{H}_{aniso}=\sum_{i<j}
\left(g_\perp (\tau^x_i \tau^x_j+\tau^y_i \tau^y_j) + g_z \tau^z_i \tau^z_j \right)
\delta^{(2)}(\mathbf{r}_i-\mathbf{r}_j),
\label{aniso}
\ee
where the $\tau^\alpha$ Pauli matrices operate in valley space. 
The model we consider is given by the total Hamiltonian $\mathcal{H}_0 +\mathcal{H}_{aniso}$.
There is a U(1) symmetry due to rotations around the isospin $z$ axis leading to conservation
of the $z$ component of isospin $T_z$. In the absence of Zeeman energy there is complete SU(2) spin symmetry which is
broken down to U(1) spin rotation around $z$ axis with generator $S^z$ for nonzero Land\'e factor.
The parameters $g_\perp,g_z$ are not precisely known but
the associated energy scale is likely to be larger than the Zeeman energy in graphene samples. 
This energy scale is nevertheless smaller than the Coulomb energy scale.
The main evidence for the role
of anisotropies is the presence of a phase transition in the conductance observed when tilting the 
applied magnetic field~\cite{Young2012,Young2014}. 
It is convenient to parameterize the two  coefficients $g_{\perp,z}$ with an angular variable $\theta$~:
\be
g_\perp = g\cos\theta , \quad g_z=g\sin\theta.
\ee
in addition to an overall strength $g$ which we take as positive.
We next define a dimensionless strength of the anisotropies by using the Coulomb energy scale as a reference point~:
\be
\tilde g = (g/\ell_B^2)/(e^2/(\epsilon\ell_B))
\ee
It is also convenient to translate the parameters $g_{\perp,z}$ into separate energy scales~:
\be
u_\perp = \frac{g_\perp}{2\pi \ell_B^2}, \quad u_z = \frac{g_z}{2\pi \ell_B^2}.
\ee
Experimental evidence suggests that the energy scale $g/\ell_B^2$ is small compared to the typical Coulomb energy
$e^2/(\epsilon\ell_B)$ so the effect of anisotropies should be seen as a small perturbation that lifts SU(4)
degeneracies. Notably it selects the $\nu=0$ ground state among several possibilities.
Let us follow the quantum Hall ferromagnetism approach~\cite{nomura,Roy2014,Khari1,Khari2} to describe the filling factor $\nu=0$. 
We have to find
two orthogonal vectors $\phi_1$ and $\phi_2$ in the four-dimensional space spanned by 
$\{|K\uparrow\rangle,|K\downarrow\rangle,|K^\prime\uparrow\rangle,|K^\prime\downarrow\rangle\}$
and fill exactly all orbital indices denoted by $m$~:
\be
|\Psi_0\rangle =\prod_m c^\dagger_{m\phi_1} c^\dagger_{m\phi_2}|0\rangle,
\label{n0wavefn}
\ee
If we neglect Landau level mixing then these states, for any choice of the pair $\phi_1,\phi_2$,
are exact eigenstates of the fully SU(4) symmetric Coulomb interaction. It is likely that they are
even exact ground states as observed numerically on small systems~\cite{Wu}. Adding anisotropies that break the full SU(4) symmetry
lifts this degeneracy and leads to the following phases studied in refs.(\onlinecite{Khari1,Khari2,Wu})~:

\begin{itemize}
 \item 
 the ferromagnetic phase F defined by $\phi_1=|K\uparrow\rangle,\phi_2=|K^\prime\uparrow\rangle$
which is stabilized in the range $-\pi/4<\theta<+\pi/2$

\item
 the antiferromagnetic phase AF with $\phi_1=|K\uparrow\rangle,\phi_2=|K^\prime\downarrow\rangle$
preferred in the range $\pi/2 < \theta  < 3\pi/4$. Note that this state is an antiferromagnet both in spin space
and in valley space.

\item
 the K\'ekul\'e state (KD) with $\phi_1=|\mathbf{n}\uparrow\rangle,\phi_2=|\mathbf{n}\downarrow\rangle$
where $\mathbf{n}$ is a vector lying in the XY plane of the Bloch sphere for valley degrees of freedom~:
$\mathbf{n}=(|K\rangle+\mathrm{e}^{i\phi}|K^\prime\rangle)/\sqrt{2}$ where $\phi$ is an arbitrary angle in the XY plane.
This state is a spin singlet but a XY valley ferromagnet.
It is preferred in the range $3\pi/4<\theta<5\pi/4$.

\item
 the charge density wave state (CDW) defined by $\phi_1=|K\uparrow\rangle,\phi_2=|K\downarrow\rangle$.
This state is a spin singlet but a valley ferromagnet. Since the valley index coincides with the sublattice index
in the central Landau level, this state has all the charge  density on one sublattice and would be favored by a substrate
breaking explicitly the sublattice symmetry like hexagonal boron nitride. It requires the range
$5\pi/4<\theta<7\pi/4$ as it is stabilized by negative valley interactions along $z$ direction.

\end{itemize}

This description is strictly valid in the absence of Zeeman energy. Notably when the Land\'e factor is not zero
the AF phase undergoes spin canting~: for small Zeeman energy it is energetically more favorable to have the spins
lying in the plane orthogonal to the direction of the field and almost antiparallel, both having a small component
aligned with the field. Increasing the Zeeman energy leads to an increased component in the field direction up to full
saturation where we recover the F phase. A mean-field calculation leads to an approximate state described by
$\phi_1=|K,\mathbf{s}_+\rangle,\phi_2=|K^\prime,\mathbf{s}_-\rangle$ with the spin states on the Bloch sphere coordinates
are given by $\mathbf{s}_\pm =(\pm \sin\alpha \cos\beta,\pm \sin\alpha \sin\beta,\cos\alpha)$ where $\beta$ is an arbitrary
angle in the XY plane of the spin Bloch sphere and $\alpha$ is the canting angle determined by the competition between
Zeeman energy and anisotropies $\cos\alpha =\epsilon_Z/(2|u_\perp|)$. This is the so-called canted antiferromagnetic phase (CAF).
Negligible Zeeman energy leads to the pure AF phase while
large Zeeman energy realized by tilting the field leads to the limit $\alpha\rightarrow\pi/2$~: the two spins become
aligned and we are in the ferromagnetic phase. The transition observed experimentally in the conductance of the samples
of ref.(\onlinecite{Young2012}) is attributed to this CAF/F transition.
Since we discuss the role of Zeeman energy in some detail we will use interchangeably AF and CAF for the antiferromagnetic phase.

There are interesting high-symmetry points in the full Hamiltonian $\mathcal{H}_0+\mathcal{H}_{aniso}$. 
They are studied in detail in ref.(\onlinecite{Wu}). Notably there
is a SO(5) symmetry at the boundary $g_\perp+g_z=0$ between the AF and the KD phases which unifies antiferromagnetism and K\'ekul\'e ordering.
In this paper we concentrate on the study of skyrmions only for AF and F phase because these are likely seen in experiments.
We note for the future that with a substrate breaking sublattice equivalence (such as hexagonal boron nitride)
it is possible to stabilize the KD phase
or the CDW phase.

In the case of the filling $\nu=-1$ the ground state wavefunction is given by~:
\be
|\Psi_{-1}\rangle =\prod_m c^\dagger_{m\phi} |0\rangle,
\label{n1wavefn}
\ee
where $\phi$ is any spinor and the sum over orbital indices $m$ fills all available states.
While this is an exact eigenstate of the SU(4) symmetric Coulomb interactions it is also an exact
eigenstate of the anisotropic Hamiltonian Eq.(\ref{aniso}) because the wavefunction Eq.(\ref{n1wavefn}) vanishes when two particles are
at the same point space and the anisotropies are taken to be purely local. It means that the degeneracy in
the choice of $\phi$ is not lifted by the simple model we use. This ambiguity persists for quasihole state that
are created from $\nu=-1$ by adding extra flux. Since the state is even less dense than Eq.(\ref{n1wavefn})
its wavefunction still vanishes when particle positions coincides. However the electron skyrmion which is more dense is sensitive to
anisotropies and its fate will be studied in section IV. A lattice model like the one studied in Ref.(\onlinecite{DNSheng})
may be necessary to pinpoint the true nature of the ground state at $\nu=-1$.
Since the anisotropies described by Eq.(\ref{aniso}) will change the nature of the electron skyrmion for $\nu=-1$
we will refer to the range $\pi/2 <\theta < 3\pi/4$ as the AF phase and $-\pi/4 < \theta < +\pi/2$ as the F phase
even for $\nu=\pm 1$.

\section{SU(2) skyrmions and the spherical geometry}
In this section we recall basic facts about SU(2) Skyrmions as studied in the spherical geometry.
We consider electrons with spin 1/2 interacting through the spin-symmetric Coulomb interaction
and no anisotropies in spin space. This means that the Hamiltonian is given by Eq.(\ref{H0}) i.e.
it is just $\mathcal{H}_0$.
If there are $N_\phi$ flux quanta through the sphere then $\nu=1$ filling of the lowest Landau level
requires $N=N_\phi+1$ electrons and in the absence of Zeeman energy the corresponding wavefunction is given
by a Slater determinant~:
\be
|\Psi_{\nu=1}\rangle = \prod_M c^\dagger_{M\chi}|0\rangle,
\label{nu1gs}
\ee
where all orbital states are occupied $M=-S,\dots, +S$ where $N_\phi=2S$ and $|\chi\rangle$ is \textit{any} spin state.
The one-body states indexed by integer or half-integer $M$ are given by~:
\be
\Phi^{(S)}_M= \sqrt{\frac{N_\phi+1}{4\pi}
\left(\begin{array}{c}
 N_\phi \\ S-M \\
 \end{array}  \right)}
 u^{S+M}v^{S-M}, \quad u=\cos \frac{\psi}{2} \mathrm{e}^{-i\eta/2}, v=\sin \frac{\psi}{2} \mathrm{e}^{i\eta/2},
\ee
where $\psi,\eta$ are spherical coordinates and the number of flux quanta through the sphere $N_\phi=2S$ is an integer.
One-body states are eigenstates of the orbital angular momentum and so many-body states can also be
classified with total angular momentum since Coulomb interactions and local anisotropies translate
into rotationally-invariant interactions on the sphere.
The state wavefunction Eq.(\ref{nu1gs}) is an orbital singlet $L_{tot}=0$ and a ferromagnetic multiplet of maximal total
spin since the state is totally spin-symmetric. From this fiducial state one can now add or remove one electron
to obtain an entity carrying charge. A Hartree-Fock approximation is simply to remove one electron from wavefunction Eq.(\ref{nu1gs})
or to add an electron with a spin state $\chi^\prime$ orthogonal to $\chi$ and in any orbital state. Indeed 
the ground states are no longer exactly given by a simple Slater determinant.

\begin{figure}[h]
\centering
\includegraphics[width=0.4\columnwidth]{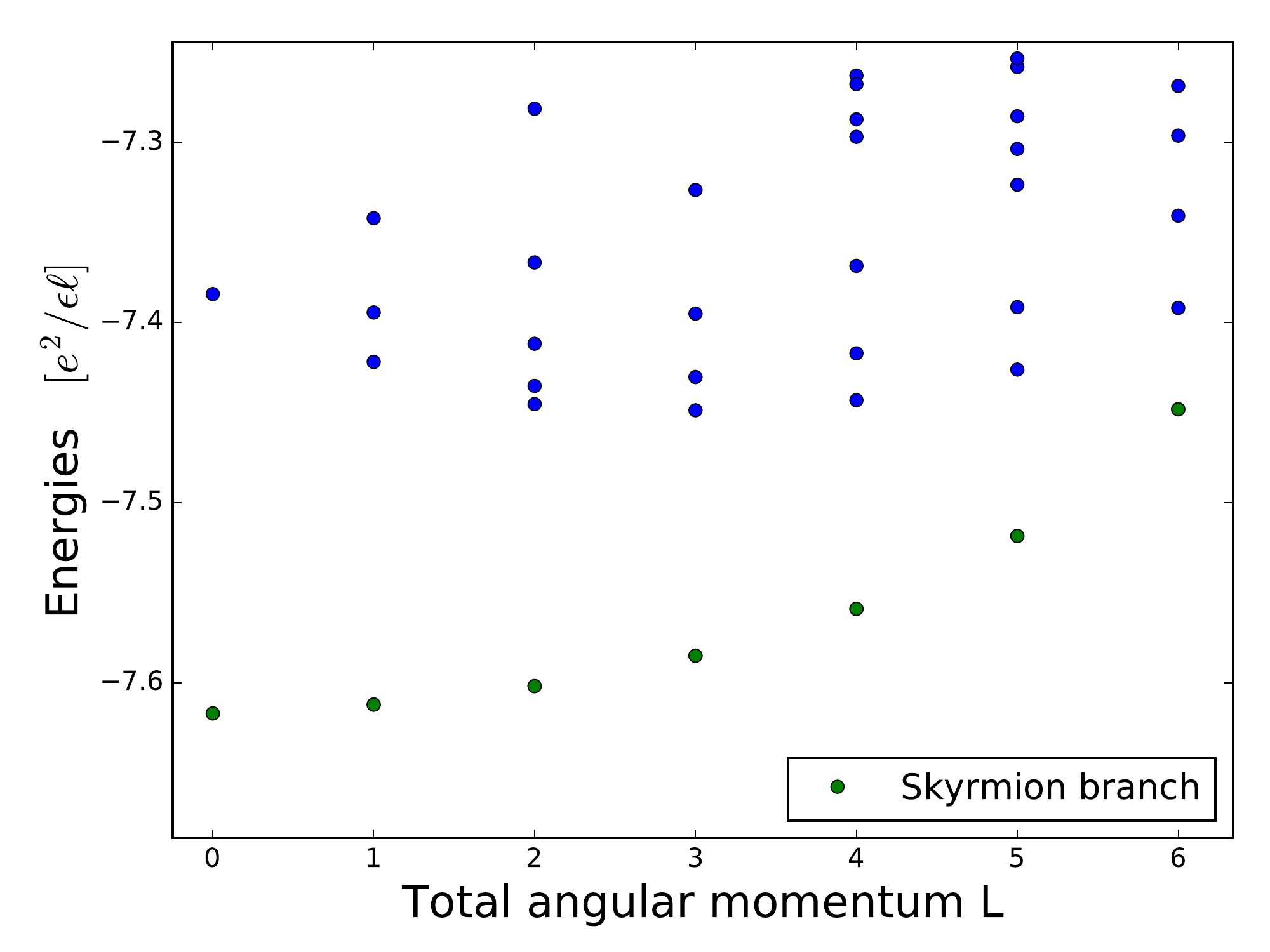}
\caption{Energy spectrum of 12 electrons on a sphere with $N_\phi=12$ flux quanta and zero Zeeman energy,
leading to the formation of a  $SU(2)$ hole skyrmion. Energies are plotted as a function of total orbital angular momentum.
There is a well-defined branch of low-lying states extending from $L=0$ up to $L=N_\phi/2$ of states
having $L=S$ that are interpreted as skyrmion states (green dots). The ground state at $L=0$ is also a spin singlet state which is
delocalized on the whole sphere. The highest-lying state  with $L=N_\phi/2=6$ of the skyrmion branch is the fully polarized hole. }
 \label{su2sk}
\end{figure}

It has been discovered numerically~\cite{rezayi91,Sondhi93} that the ground state becomes a spin and orbital singlet when 
changing by one unit the number of charges in the system. In addition to a $L_{tot}=S_{tot}=0$ ground state there is a low-lying well-defined 
branch of states that have equal spin and orbital momentum $L=S$ that rises up to $L=S=N_\phi/2$. It is well isolated
from higher-lying states~: see Fig.(\ref{su2sk}) for a typical spectrum in the absence of Zeeman energy. 
The end of this skyrmion branch is reached for $L=N_\phi/2$
by the fully polarized hole state whose exact wavefunction can be obtained by removing one creation operator
in the wavefunction Eq.(\ref{nu1gs}). 
This skyrmion branch is the finite-size translation of the skyrmion spin texture described by the nonlinear sigma model~\cite{Moon95,Yang96}.
The shape of the branch is a reflection of interaction potential between electrons.
Indeed by using a hard-core interaction between same-spin electrons it becomes perfectly flat.

It is convenient to introduce the number of overturned spins $K$ with respect
to the fully polarized case by $L=S=N_\phi/2-K$~: $K$ increases monotonically as we go along the skyrmion branch starting 
from the polarized hole with $K=0$. The ground state with $L=0$ is fully delocalized on the sphere and corresponds to a skyrmion of
infinite size in the thermodynamic limit. It is reached for larger and larger $K$ as the number of particles grows.
The exponential growth of the Fock space with the number of particles severely limits the sizes that can be reached
by exact diagonalization. It means we can study only skyrmions with a small number of overturned spins. To overcome
this problem it is also possible to use trial wavefunctions in Hartree-Fock calculations.
They can written in the disk geometry as follows~:
\be
|\Psi_h\rangle = \prod_{m=0}^\infty (u_m c^\dagger_{m\uparrow} 
+ v_m c^\dagger_{m+1\downarrow})|0\rangle
\label{Dsk}
\ee
where the one-body states indexed by a positive integer $m$ are the symmetric gauge lowest Landau level eigenstates~:
\be
\phi_m(z)=\frac{z^m}{\sqrt{2\pi 2^m m!}}\exp(-|z|^2/4\ell_B^2),
\label{Landau}
\ee
where we have defined the complex planar coordinate $z=x+iy$ and $m\geq0$ is a positive integer.
The coefficients $u_m$ and $v_m$ are variational parameters.

\begin{figure}
\centering
\includegraphics[width=0.4\columnwidth]{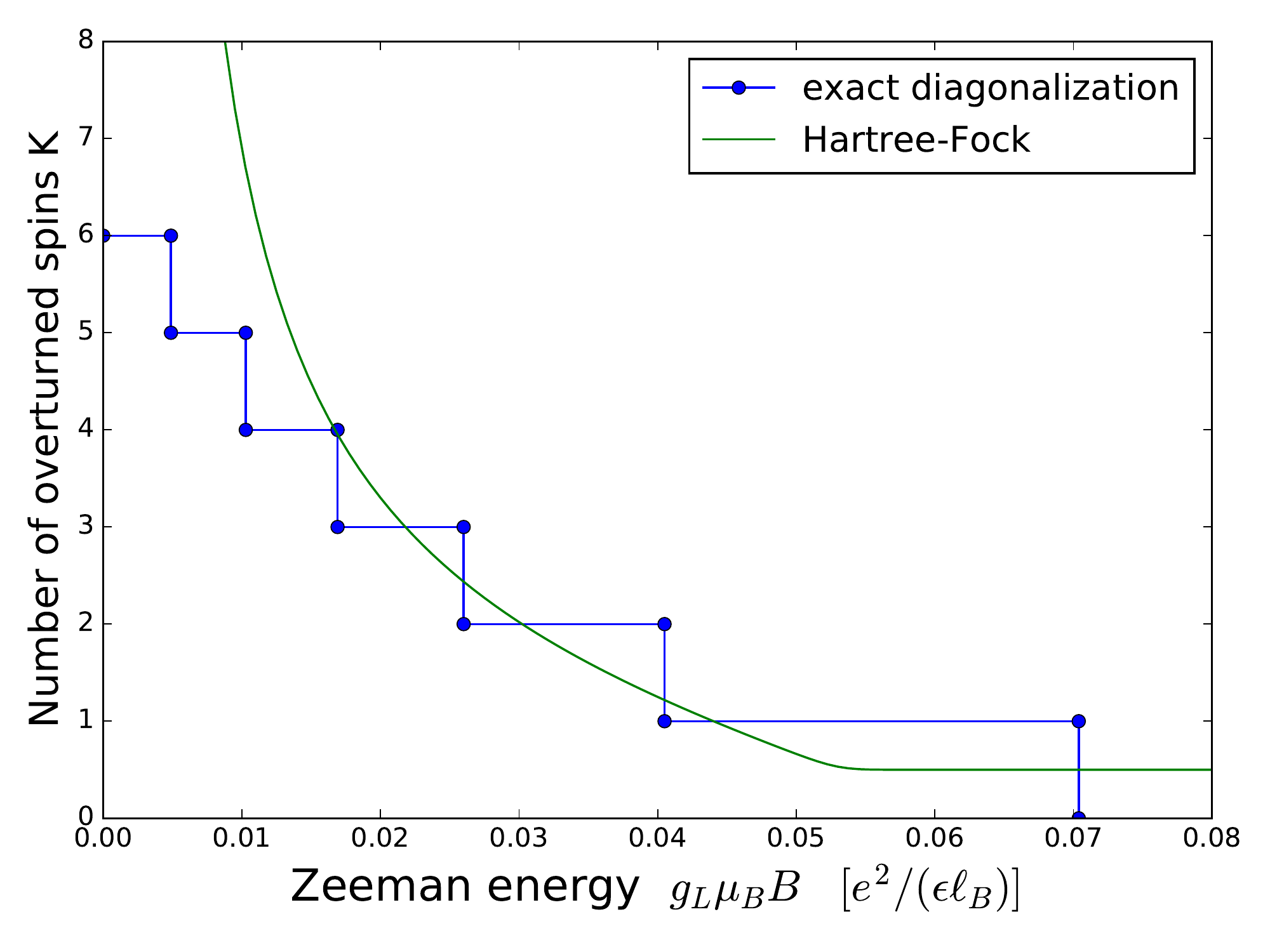}
\caption{The number of overturned spins in a SU(2) hole skyrmion as a function of Zeeman energy. 
The blue steps indicate the level crossings as observed by exact diagonalization of $N=12$ electrons on a sphere
with $N_\phi=12$. The green curve comes from a Hartree-Fock calculation with the variational state Eq.(\ref{Dsk}).
A finite Zeeman energy is required to fully polarize the system.}
 \label{magsu2}
\end{figure}

The magnetization process of the skyrmion involves both spin and orbital angular momentum. In the absence of the Zeeman
term the ground state is the $L=0$ member of the skyrmion branch. This describes a state with no characteristic length
scale it is delocalized over all the sphere and corresponds to a skyrmion of infinite size. With finite Zeeman coupling all higher-lying
states along the skyrmion branch are split according to their spin value~: since for a finite system there are finite energy
separations between these states there will be a succession of level crossings when states with increasing spin become the ground state
in the presence of the Zeeman effect. The magnetization curve is then a staircase  as a function of the applied field $B$.
The critical fields are defined through~:
\be
g_L\mu_B B_{crit}^{(N/2-K)}=E_{L=K+1}-E_{L=K}.
\ee
See e.g. Fig.(\ref{magsu2}) where we plot the number of overturned spins $K$ with respect to the fully saturated case.
It is only beyond some critical field that the ground state becomes fully magnetized when a state from the multiplet with $L=N_\phi/2$
becomes the ground state. If we apply the Hartree-Fock (HF) method to the state Eq.(\ref{Dsk}) along the lines of 
refs.(\onlinecite{fertig,Fertig97,Brey96,BreyL,mfb})
we obtain another approximation to the skyrmion state and thus of the magnetization curve. In Fig.(\ref{magsu2}) we have plotted
the HF result in addition to the curve extracted from exact diagonalization with 12 electrons on a sphere.
We recover the typical skyrmion spin flip of $K\approx 3$ when $g_L\mu_B B\approx 0.015 e^2/(\epsilon\ell_B)$
in GaAs systems.
The various approaches to the skyrmion properties have been compared in ref.(\onlinecite{Abolfath97}).
It is important to note that the Coulomb interaction is responsible for the shape of the skyrmion branch
and there is a competition between this effect and the Zeeman energy to determine the skyrmion profile.
This simple picture will break down in the SU(4) case as we discuss in the following.

\section{Skyrmions at $\nu =\pm 1$}
If we first neglect anisotropies we can make exact statements about SU(4) skyrmions at filling factor $\nu=-1$
and its particle-hole partner filling $\nu=+1$. Indeed if we consider exact eigenstates of the SU(2) symmetric
situation without Zeeman coupling, these states can be embedded in the larger Fock space built for four flavors
and since Coulomb interactions are fully SU(4) symmetric these states will remain exact eigenstates. The only
characteristic that changes is the degeneracy since one spin flip is exactly equivalent to any flavor flip, only
the counting of states is different. So we know that the SU(2) skyrmion states are still present in the SU(4) case.
The only question is whether or not intruder states will destroy this picture. This question can be studied
by our exact diagonalizations and the answer is simple~: there are no new intruder states with SU(4) character
beyond the states we know from the SU(2) case. We find that this is valid for the whole skyrmion branch~: the picture corresponding
to our Fig.(\ref{su2sk}) is thus exactly the same in the SU(4) case. Only the degeneracies of the states change.
There are of course additional new states
but they lie at higher energies in the upper part of the spectrum. This is evidence for the correctness of the discussion given in 
ref.(\onlinecite{Arovas99}). These arguments equally apply to $\nu=\pm 1$ and $\nu=0$. 

According to refs.(\onlinecite{Arovas99,Yang2006}) the degeneracies can be computed straightforwardly with Young tableaux.
The skyrmion state with largest angular momentum spans an SU(4) irreducible representation (IR) given by one row
of $N_\phi$ boxes~:
\begin{equation}
L=N_\phi/2~:\quad
\begin{ytableau}
~ & ~ & \dots & \dots & \dots & ~ &  ~ \\ 
\end{ytableau}
\end{equation}
then the neighboring state with $K=1$ has one row of $N_\phi-1$ boxes and one row of one box~:
\begin{equation}
L=N_\phi/2-1~:\quad
\begin{ytableau}
~ & ~ & \dots & \dots & ~ &  ~ \\ 
~
\end{ytableau}
\end{equation}
We move then along the whole skyrmion branch by removing boxes from the upper row and transferring them
to the second row~:
\begin{equation}
L=N_\phi/2-2~:\quad
\begin{ytableau}
~ & ~ & \dots & \dots & ~ &  ~ \\ 
~ & ~
\end{ytableau}
\end{equation}
we proceed
till there are $N_\phi/2$ boxes in both rows (we assume $N_\phi$ even for simplicity, otherwise
there is one remaining box).
\begin{equation}
L=0~:\quad
\begin{ytableau}
~ & ~ & \dots & ~ &  ~ \\ 
~ & ~ & \dots & ~ & ~
\end{ytableau}
\end{equation}
In the SU(2) case we can omit the second row and we just obtain the spin multiplets with $L=S$.
In the SU(4) case these IRs have dimensions given by~:
\be
\mathcal{D}(a,b)=\frac{1}{2}(a+1)(b+1)(b+2)(a+b+2)(a+b+3),
\ee
where $b$ is the number of columns with two boxes and $a$ is the number of columns with one box.

\begin{figure}
\centering
\includegraphics[width=0.4\columnwidth]{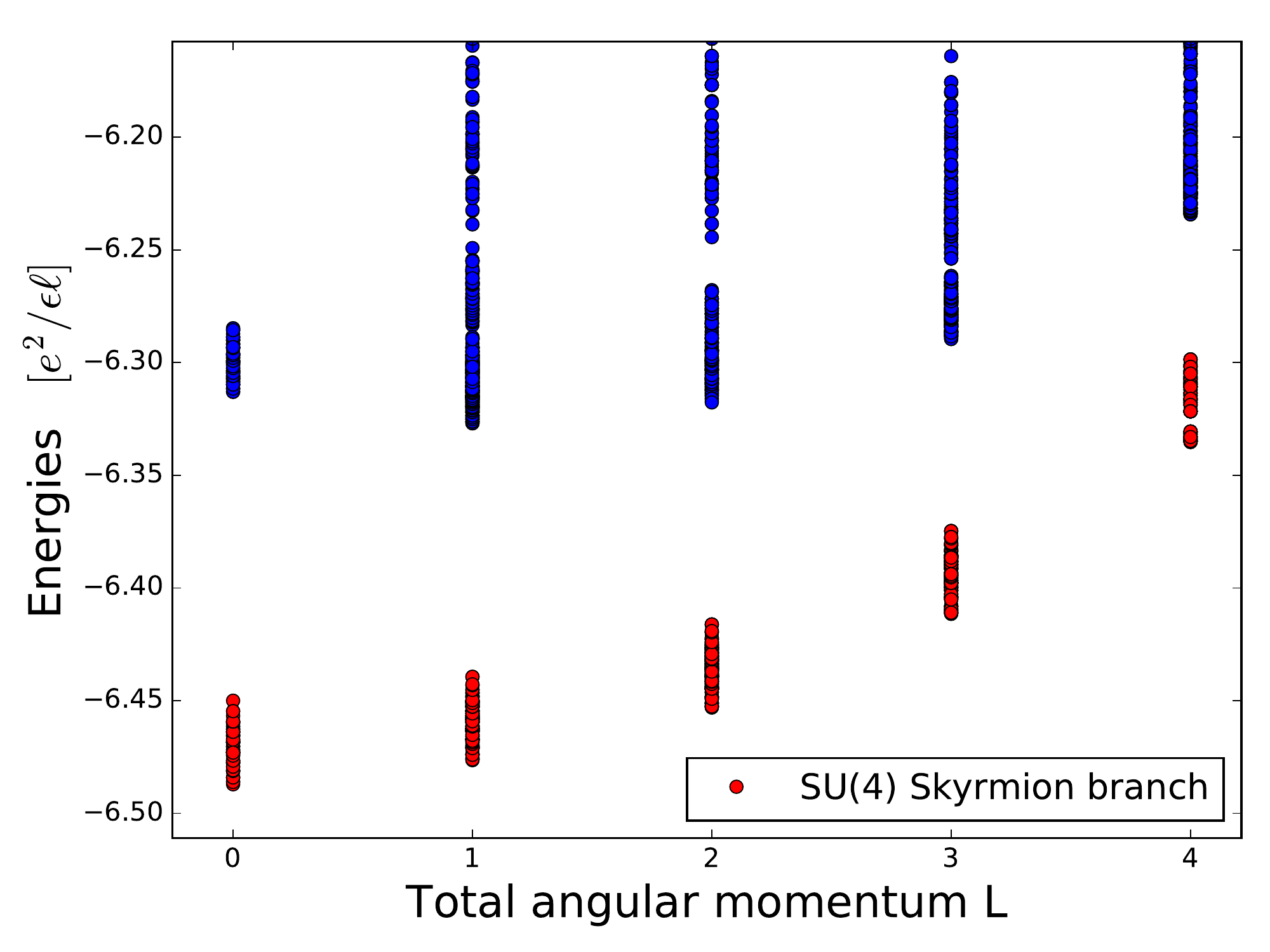}
\caption{Energy spectrum of 10 electrons on a sphere with $N_\phi=8$ flux quanta and zero Zeeman energy.
There is formation of a $\nu=-1$ electron skyrmion. Energies are plotted as a function of the total angular momentum.
The anisotropies are taken to be $\tilde g=0.1$ and the anisotropy angle is $\theta=0.6\pi$ so the system is in the AF phase. 
All SU(4) multiplets
along the skyrmion branch are now split into manifolds of states with an energy splitting scale $O(\tilde g)$ 
(even with a very large anisotropy $O(1)$
the skyrmion picture is not destroyed). The value $\tilde g=0.1$ is chosen for readability of the figure, realistic values are likely
smaller.}
 \label{skbranch}
\end{figure}

We now turn to the role of
graphene-specific anisotropies at $\nu=\pm 1$. 
In the absence of the Zeeman term, eigenstates are classified by total angular momentum and spin as in the SU(2) case
but the anisotropic Hamiltonian Eq.(\ref{aniso}) allows conservation of valley isospin along $z$ direction.
The corresponding isospin $T_z$ is a good quantum number that we use in our exact diagonalizations in addition to
$L_z$ and $S_z$. This is for a generic angle~: special values of $\theta$ in Eq.(\ref{aniso})
have more conserved quantities which we don't consider here.
As long as the magnitude $\tilde g$ of the anisotropy remains small each state of the skyrmion will have its SU(4) degeneracy lifted
but the overall picture remains valid. In Fig.(\ref{skbranch}) we have plotted the spectrum of a 10-electron system with flux tuned
to create an electron skyrmion. The states of the skyrmion branch are now split in a series of multiplets and for small enough
anisotropy they still keep the shape we associate to skyrmion states from the SU(2) case. Even for an unrealistically large value of
$g/\ell_B^2=e^2/(\epsilon\ell_B)$ the skyrmion survives. The effect of anisotropies of Eq.(\ref{aniso}) for small $\tilde g$ is simply
 a first-order perturbation theory lifting the degeneracies between members of a given $L$ multiplet.
The eigenstates with nonzero anisotropies still have  good quantum numbers~: total spin and also $T_z$. But now we have to take into
account the fact the magnetization process that takes place when we increase the Zeeman energy is quite different from
what happens in the SU(2) discussed in section III. Indeed for each multiplet along the skyrmion branch there are states with
all possible values of the spin form $S=0$ up to $S=N/2$. So each multiplet has its own magnetization curve and thus one has to
figure out which multiplet will describe the magnetization process in the thermodynamic limit. 
An important consequence is that the energy scale of the magnetization curve is set by the anisotropy energies and not by the Coulomb scale~:
\be
g_L\mu_B B_{crit}^{(S)}=E_{S+1}-E_{S}=O(g/\ell_B^2).
\label{mag-univ}
\ee
This is due to the fact that all level crossings corresponding to the magnetization process occur within a manifold of states
whose degeneracy is lifted solely by non-zero anisotropies.

We now discuss the level ordering in the two phases of interest, AF and F. The situation in the F phase is very simple indeed~:
for each value of $L$ the multiplet has a ground state with the maximal spin $S=N/2$ as can be guessed by the ferromagnetic nature
of the anisotropies. As a consequence the maximal spin state with $L=0$ will be the ground state for \textit{all} values of the Zeeman energy.
This means that in the F phase the skyrmions are always of infinite size and always fully polarized for all values of the Zeeman field.
This is very different form from the case of SU(2) skyrmions where the fully polarized quasiparticle is the state which is maximally localized
with a size of order $\ell_B$. Concerning the valley degrees of freedom the ground state has always the smallest possible value
of $T_z$ i.e. $T_z=0$ for N even and $T_z=1/2$ for N odd. So the skyrmion is valley unpolarized.

We have checked that this picture is valid deep in the F phase but we cannot track the eigenstates close to
the transition point $\theta=\pi/2$ because there are additional degeneracies due to the enhanced symmetry rendering delicate the convergence
of Krylov-subspace methods.

\begin{figure}[h]
\centering
\includegraphics[width=0.4\columnwidth]{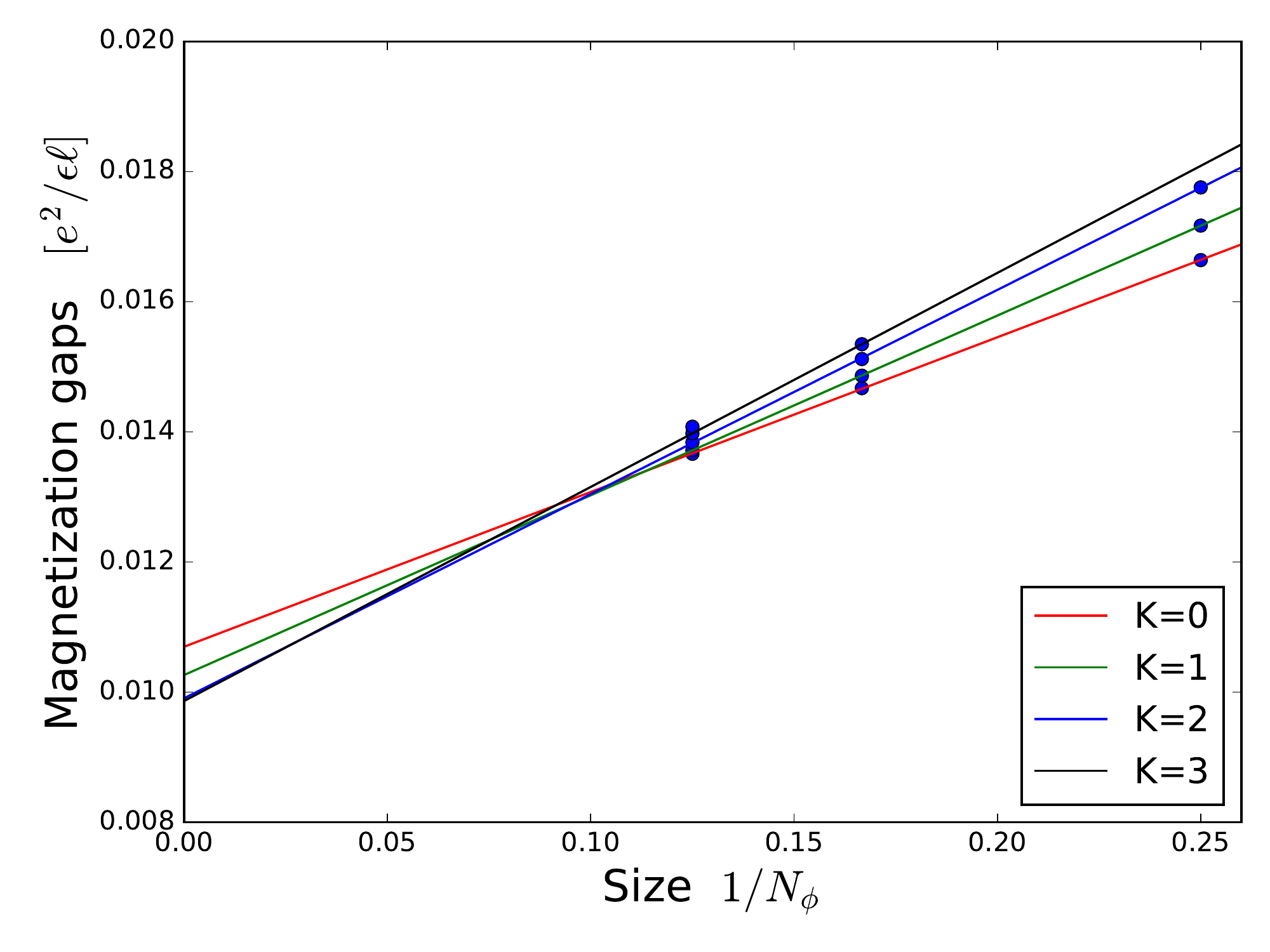}
\caption{The gaps between the singlet state and the fully magnetized state for a given $K$ value as a function of
system size. The values are computed in the AF phase with $\theta=0.6\pi$ and $\tilde g=0.1$.
The tentative linear fits given by the straight lines are listed in Table \ref{Kgaps}.}
 \label{maggaps}
\end{figure}

\begin{table}[ht]
\begin{center}
\resizebox{4cm}{!}{
\begin{tabular}{|c|c|}\hline
 K & magnetization gap $G(K)$ \\  \hline
 0  &  0.0107 \\
 1  &  0.0102 \\
 2  &  0.0099 \\
 3  &  0.00985\\
 \hline
\end{tabular}
}
\end{center}
\caption{Estimates of the gaps between the spin singlet ground state at fixed $K$ value and the lowest-lying
fully spin polarized state. Gaps are given in units of the Coulomb scale $e^2/(\epsilon\ell_B)$ and
parameters are the same in Fig.(\ref{maggaps}): $\theta=0.6\pi$ and $\tilde g=0.1$.}
\label{Kgaps}
\end{table}

\begin{figure}[h]
\centering
\includegraphics[width=0.4\columnwidth]{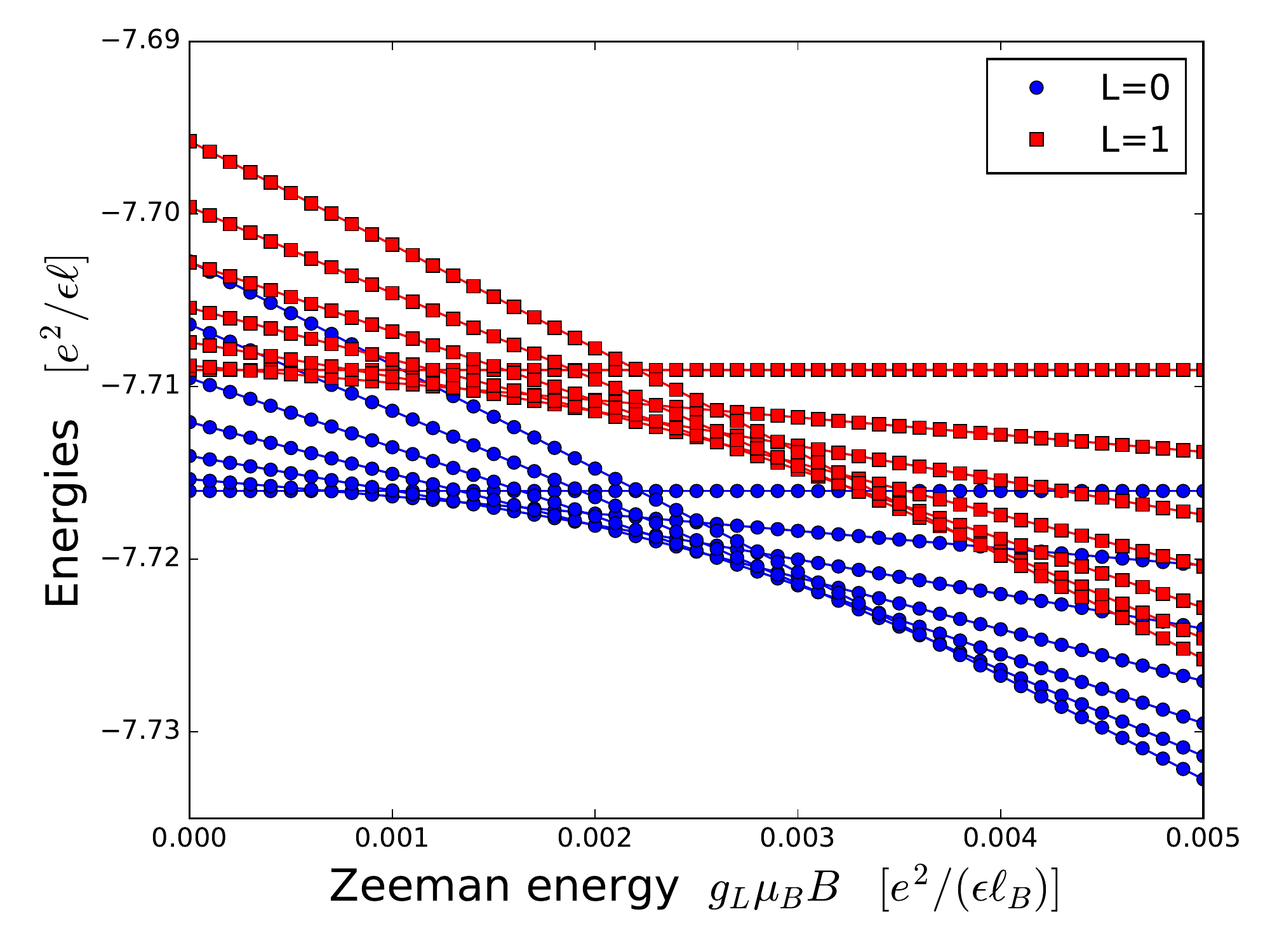}
\caption{Evolution of energy levels of a 10-electron system with an electron skyrmion due to Zeeman effect~:
we have plotted only levels with spin ranging from $S=0$ up to the maximal value $S=N/2$
and for clarity followed only the levels with largest $S^z$ value of each multiplet~:
these are the states that have the largest slope as a function of $B$. Blue traces belong to the multiplet of states with
$L=0$ while red traces are coming from the $L=1$ multiplet. So the magnetization is always entirely due to the $L=0$ states.
The envelope of these states determines the magnetization process of the skyrmion.
Our finite-size studies show that this is still the case in the thermodynamic limit.}
 \label{mag01}
\end{figure}

\begin{figure}
\centering
\includegraphics[width=0.4\columnwidth]{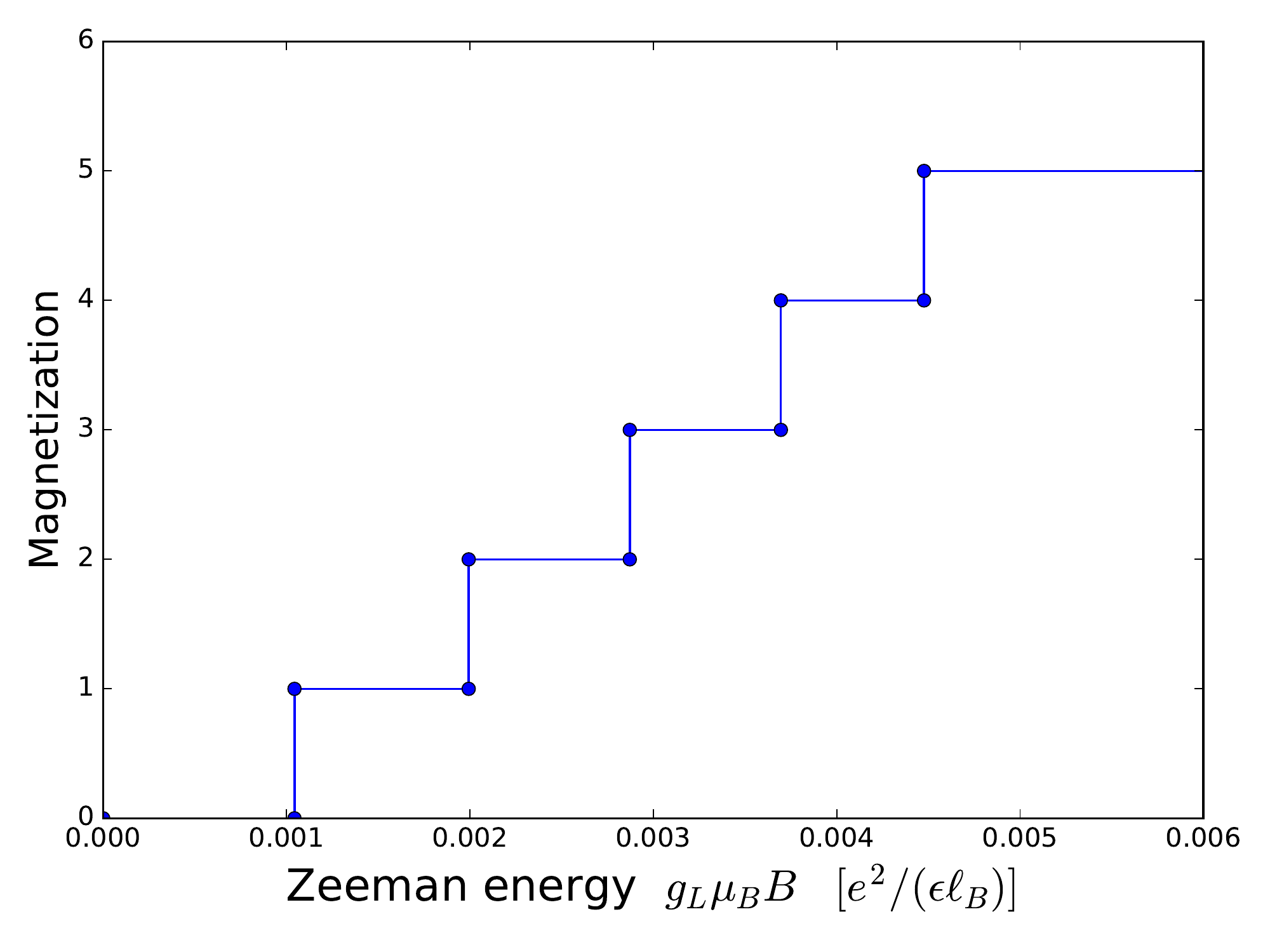}
\caption{The magnetization of the set of $L=0$ states for the electron skyrmion at $N_\phi=8$ in the AF
phase with $\theta=0.6\pi$ and $\tilde g=0.1$. Note that this staircase curve will collapse onto the vertical axis when
we reach the F phase for $\theta=0.5\pi$.}
 \label{magnetcurveNphi8}
\end{figure}

In the AF phase we observe now that the ordering of the multiplets is reversed with respect to the F case and
now the ground state at fixed $L$ has always zero spin. The magnetized states have energies increasing with magnetization  at fixed $L$.
It is convenient to index states along the skyrmion branch again by $K=N_\phi/2-L$, $K=0,1,\dots$. For each $K$ we 
define the magnetization gap
$G_{N_\phi}(K)$ as the difference in energy between the ground state at this $K$ and the lowest-lying state which is fully polarized
and hence has a spin $S=N/2$, however this spin value is now unrelated to $K$ contrary to SU(2) skyrmions.
These gaps are plotted in Fig.(\ref{maggaps}) as a function of $N_\phi=4,6,8$ for the accessible values of $K$.
They converge to a nonzero value in the thermodynamic limit $G(K)$ which we estimate by linear fits in Table (I).
They are expected to be $O(\tilde g)$ for small $\tilde g$. We observe that $G(K)$ is decreasing monotonously 
towards a nonzero value with increasing $K$.
This means that if we consider the effect of the Zeeman splitting  the lowest levels will always be those
from the $L=0$ set of states. In Fig.(\ref{mag01}) the Zeeman splitting of the $L=0$ and $L=1$ states of the largest system we
could reach with 12 electrons is displayed. 
We have taken into account only the lowest lying states with increasing spin and added a Zeeman coupling only to states
whose energy decreases with the field for clarity.
The lower envelope of the set of curves is entirely due to the states coming from 
the $L=0$ multiplets.
It is precisely this envelope that determines the magnetization curve.
So the magnetization process is entirely due to the $L=0$ multiplet. We plot
in Fig.(\ref{magnetcurveNphi8}) the magnetization process for a system of 10 electrons at $N_\phi=8$.
All energy scales are proportional to $\tilde g$ for small $\tilde g$ so with data for $N_\phi=4,6,8$ this means 
that the critical Zeeman energy to fully saturate the system is $\approx 0.05\tilde g$ in units of the Coulomb energy.
It is important to note that due to the special magnetization process as deduced from Eq.(\ref{mag-univ})
the magnetization curve is proportional to $\tilde g$ (as long as $\tilde g\lesssim 1$) so it is 
universal with respect to the magnitude of the anisotropies. Of course it still depens upon the value of $\theta$.
The AF skyrmion is thus of infinite size, is valley unpolarized because all relevant states have also the minimal allowed
value of $T_z$ and requires a critical Zeeman energy to be fully polarized~:
\be
g_L \mu_B B_{crit}\approx 0.05 (g/\ell_B^2).
\label{criticalfield1}
\ee

\section{Skyrmions at $\nu=0$}

\begin{figure}
\centering
\includegraphics[width=0.4\columnwidth]{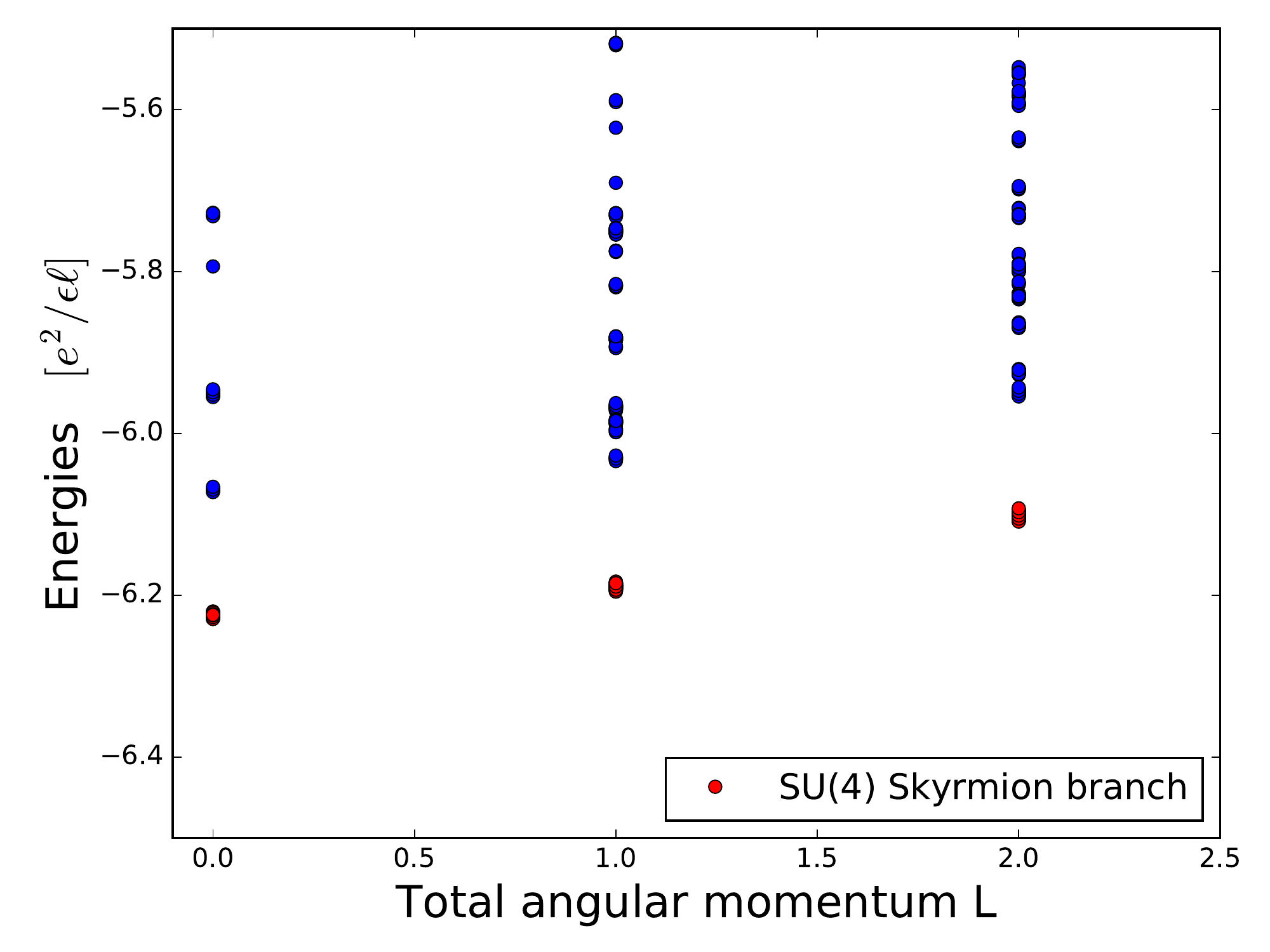}
\caption{The $\nu=0$ case~: Energy spectrum for 9 electrons at $N_\phi=4$ on a sphere leading to one hole skyrmion.
The corresponding quantum states have $L=0,1,2$ marked by red dots. The degeneracies are those predicted
by standard skyrmion counting extended to four flavors.
The eigenvalues are computed in the AF phase with $\theta=0.6\pi$ and $\tilde g=0.01$ (chosen for readability).
The important observation is that there are no new states that change the overall picture. All new physics is
entirely contained in the multiplet structure of the skyrmion branch.
This observation is valid for all values of $\theta$ in the F and AF phases we explored and up to $\tilde g\lesssim 1$.}
 \label{nu0sk}
\end{figure}

\begin{figure}
\centering
\includegraphics[width=0.4\columnwidth]{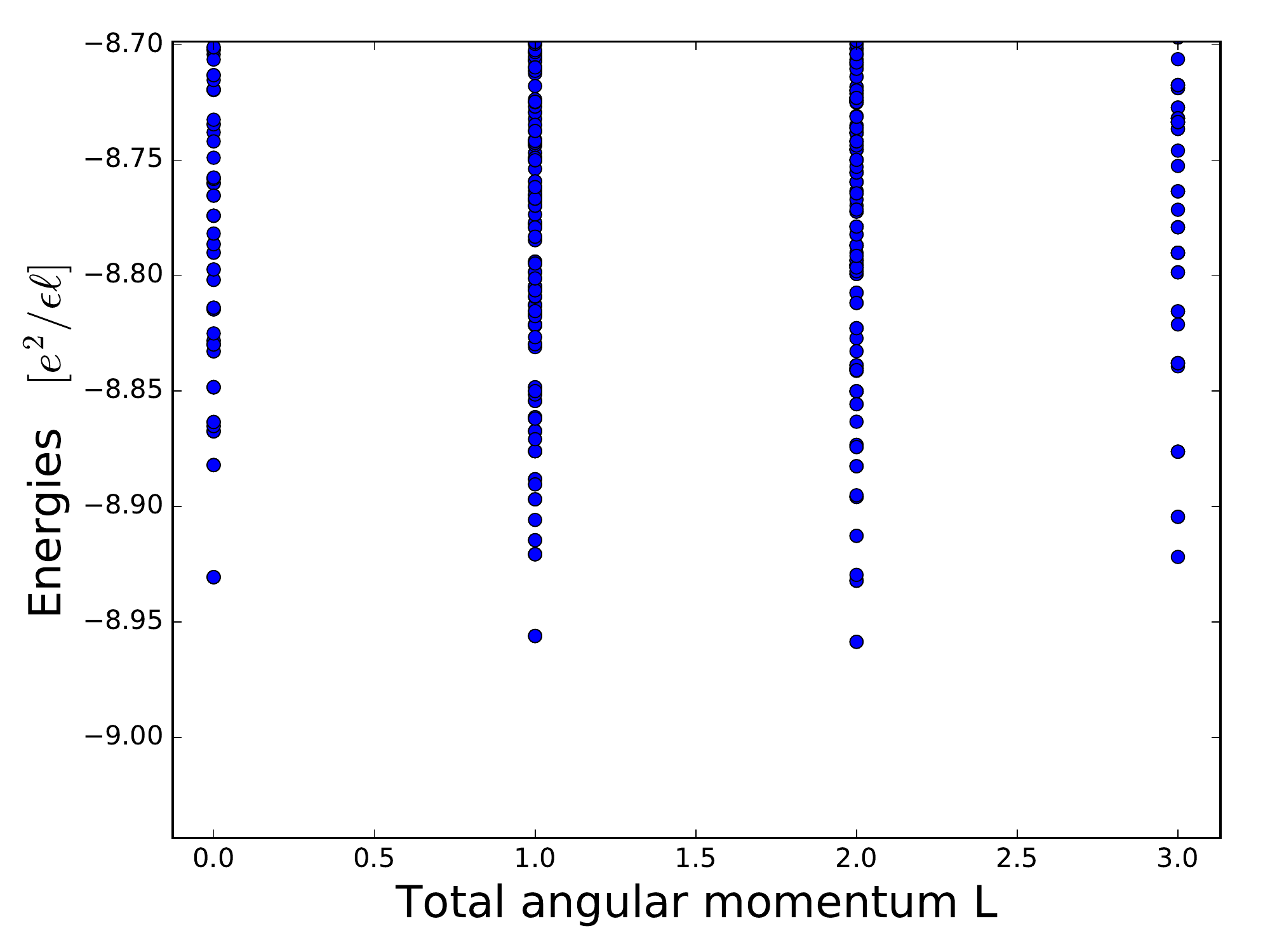}
\caption{Lower part of the energy spectrum for 13 electrons at $N_\phi=6$ on a sphere leading to one hole skyrmion.
The skyrmion states have $L=0,1,2,3$.
The eigenvalues are computed in the AF phase with $\theta=0.6\pi$ and $\tilde g=0.3$.
The ground state has no longer $L=0$. Indeed with this choice of anisotropy and size it has orbital angular momentum $L=2$ meaning that
the skyrmion has now a finite size instead of spreading all over the sphere.
This is in the absence of a Zeeman field.}
 \label{L2gs}
\end{figure}

\begin{figure}
\centering
\includegraphics[width=0.4\columnwidth]{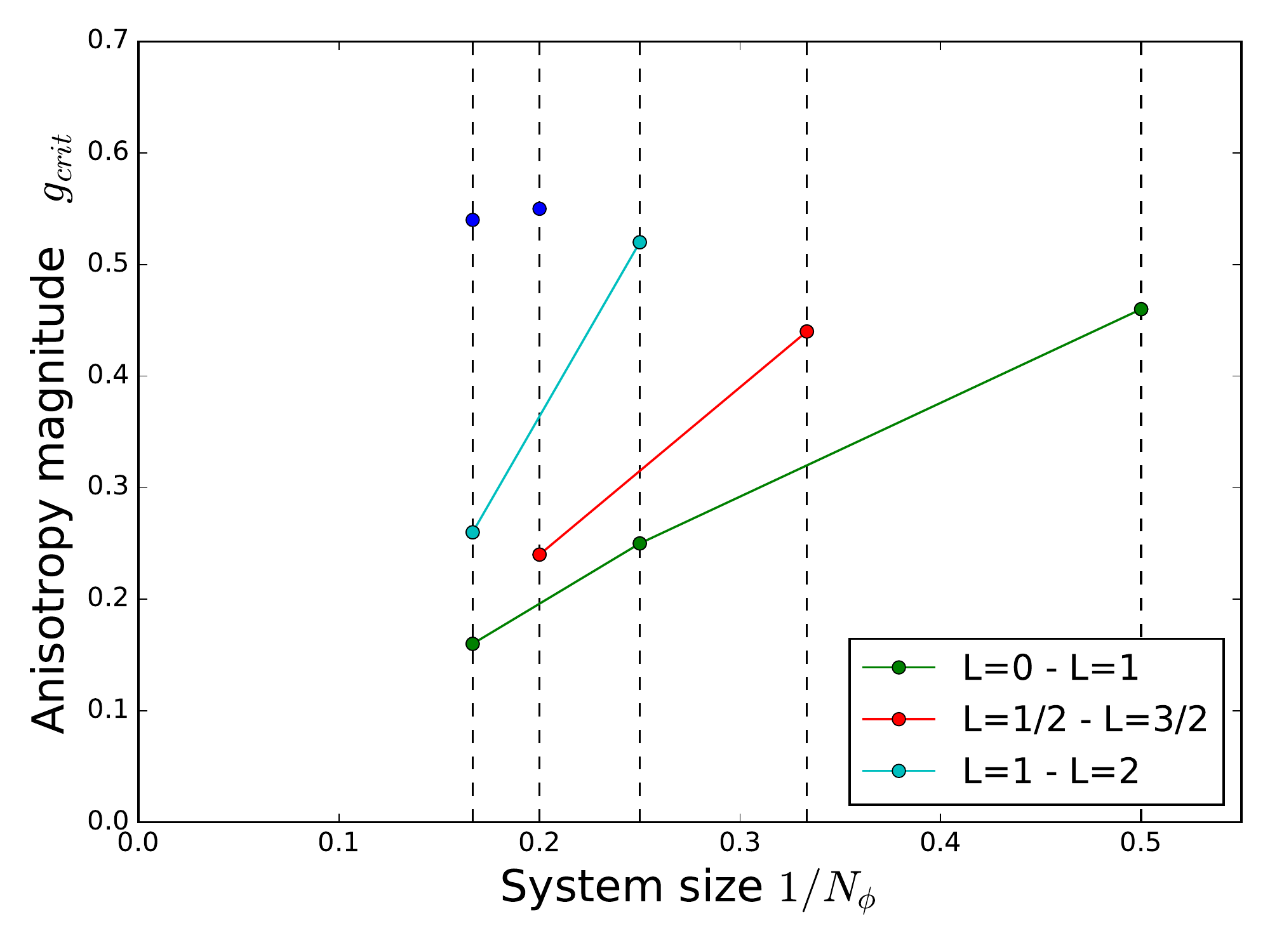}
\caption{The critical anisotropies separating ground state with distinct orbital angular momentum on the spherical geometry
as a function of system size. Below the green line systems with $N_\phi=2,4,6$ have a $L=0$ ground state and $L=1$ above.
The blue line marks the separation between $L=1$ and $L=2$ for $N_\phi=4,6$. System with odd $N_\phi=3,5$ have half-integer
orbital angular momentum and have the minimal value $L=1/2$ below the red line and $L=3/2$ above. Extrapolation of these
boundaries are all zero. At fixed anisotropy $\tilde g$ this means that a large enough system has a ground state
with maximal $L$, the end  point of the skyrmion branch. Calculations were performed in the AF phase with $\theta=0.6\pi$
and similar results hold for $\theta=0.2\pi$ in the F phase.}
 \label{gsl}
\end{figure}

We now discuss the fate of the skyrmion excitations at filling factor $\nu=0$. 
The ground state at $\nu=0$ is described by the wavefunction
Eq.(\ref{n0wavefn}). The corresponding SU(4) IR is described by a Young tableau with two rows of equal length
(when $N_\phi$ is even). We consider now only the case of the hole skyrmion because for $\nu=0$ it is related to 
the electron skyrmion by particle-hole symmetry.
If we want to create a skyrmion the recipe proposed by Ref.(\onlinecite{Yang2006}) is to glue an additional
 row of boxes describing a full inert shell on top of the Young tableaux describing the skyrmion branch of $\nu=\pm 1$. 
 So the member of the branch with maximal $L$
belongs to the IR~:
\begin{equation}
L=N_\phi/2~:\quad
\begin{ytableau}
~ & ~ & \dots & \dots & ~ &  ~ & ~\\
~ & ~ & \dots & \dots & ~ & ~ \\ 
\end{ytableau}
\end{equation}
with the $N_\phi+1$ boxes in the top row and $N_\phi$ boxes in the lower row.
We now move along the skyrmion branch with decreasing $L$ by moving boxes from the second row to a third row~:
\be
L=N_\phi/2-1~:\quad
\begin{ytableau}
~ & ~ & \dots & \dots & ~ &  ~  & ~\\
~ & ~ & \dots & \dots & ~  \\ 
~ \\
\end{ytableau}
\ee
For a state with $L=N_\phi-K$ the top row always has $N_\phi+1$ boxes the second row has $N_\phi-K$ boxes
and the third row $K$ boxes.
We have checked that the multiplicities of the states we find from our ED calculations are exactly reproduced
by the dimension of these IRs. 
An example spectrum is given in Fig.(\ref{nu0sk}).
As in the $\nu=\pm 1$ case addition of anisotropies lift the degeneracies and does not
blur the overall picture for $\tilde g$ up to unity. However we observe a new phenomenon~: the ground state of the skyrmion branch does
not stay at $L=0$ even in the absence of Zeeman energy~: see Fig.(\ref{L2gs}). In fact increasing the strength $\tilde g$ shifts the ground state 
to larger and larger values of $L$ at fixed
size $N_\phi$. This means that the skyrmion has no longer a finite size due to anisotropies in line with the arguments presented
in ref.(\onlinecite{Abanin2013}). For $\tilde g$ large enough the ground state has the smallest allowed size of order $\ell_B$ for the maximum
$L$ of the branch. In table (\ref{Nphi4QN}) and table (\ref{Nphi6QN}) we give the ground state quantum numbers as a function of $\tilde g$. 
This phenomenon happens 
both in the F and the AF phase. If we plot the critical anisotropies as a function of system size
we observe that fixing $\tilde g$ and increasing the system size leads always to the ground state with maximal $L$ value
given by $N_\phi/2$ : see Fig.(\ref{gsl}). Moving towards large system size always leads to a skyrmion state with the maximal
value of $L$. In the case of the SU(2) skyrmion that would mean complete polarization. Here it is not the case. Indeed the value
of $K$ is now decoupled from the spin value which is determined by the level ordering inside the IR with this maximal $L$ value.

\begin{table}[ht]
\begin{center}
\resizebox{4cm}{!}{
\begin{tabular}{|c|c|c|c|c|c|c|}\hline
 $\tilde g$  &  0.1 & 0.2 & 0.3 & 0.4 & 0.5 & 0.6 \\ \hline
 L  &  0  &   0 &  1  &  1  & 1  &  2\\ 
 \hline
\end{tabular}
}
\end{center}
\caption{The angular momentum of the ground state of a skyrmion state with $N_\phi=4$ and $N=9$ electrons
as a function of anisotropy scale $\tilde g$ for $\theta=0.6\pi$.}
\label{Nphi4QN}
\end{table}

\begin{table}[ht]
\begin{center}
\resizebox{4cm}{!}{
\begin{tabular}{|c|c|c|c|c|c|c|}\hline
 $\tilde g$  &  0.1 & 0.2 & 0.3 & 0.4 & 0.5 & 0.6 \\ \hline
 L  &  0  & 1 &  2  &  2  & 2  &  3\\ 
 \hline
\end{tabular}
}
\end{center}
\caption{The angular momentum of the ground state of a skyrmion state with $N_\phi=6$ and $N=13$ electrons
as a function of anisotropy scale $\tilde g$ for $\theta=0.6\pi$.
If one increases the system size at fixed anisotropy the ground state ends up at $L=0$.}
\label{Nphi6QN}
\end{table}

So at fixed $\tilde g$ we observe that the system when it is large enough has a localized skyrmion excitation. The magnetization
process that takes place now is similar to the case of $\nu=\pm 1$. It is the Zeeman splitting of the multiplet
$L=N_\phi/2$ that governs the magnetization. There we find the difference between F and AF phase. In the F case the
lowest-lying state at $L=N_\phi/2$ has maximal spin so the system is always fully polarized~: charged excitations are
fully polarized quasiparticles. On the contrary in the AF phase there is progressive magnetization with increasing Zeeman energy.
It is important to note that due to the multicomponent nature of the system the spatial and the magnetization are
no longer coupled, contrary to spin SU(2) skyrmions. To fully spin polarize the skyrmion we estimate that the Zeeman energy should be~:
\be
g_L\mu_B B_{sat}\approx 0.16\,\tilde g \times (e^2/(\epsilon\ell_B))=0.16\,(g/\ell_B^2),
\label{criticalfield0}
\ee
for $\theta =0.6\pi$. This value is approximately three times higher than in the case of the $\nu=-1$ electron skyrmion.

\section{Conclusions}

We have performed a study of skyrmion physics in monolayer graphene for Landau level filling factors
$\nu=\pm 1,0$ by use of exact diagonalization on the spherical geometry. The lattice scale anisotropies that break
the SU(4) spin/valley symmetry have been incorporated by a simple local effective interaction with two parameters.
The skyrmion physics for $\nu=\pm 1$ on one side and $\nu=0$ on the other side has the specific feature that the ground state
angular momentum on the sphere is not related to the spin polarization contrary to ordinary SU(2) skyrmions.
We have shown that state counting from straightforward application of SU(N=4) skyrmion theory is correct, i.e.
there are no additional states in the low-energy spectrum.

\begin{table}[ht]
\begin{center}
\resizebox{12cm}{!}{
\begin{tabular}{|c|c|c|}\hline
  & AF & F   \\  \hline
  \multirow{2}{*}{$\nu=-1$}
  &  hole-sk~: infinite-sized valley, fully spin-polarized & hole-sk~: inifinite-sized valley, fully spin-polarized \\ \cline{2-3}
  & e-sk~: infinite-sized, partially spin-polarized & e-sk~: infinite-sized, fully spin-polarized,  \\ \hline
 $\nu=0$  & localized, partially spin-polarized & localized, fully spin-polarized \\
 \hline
\end{tabular}
}
\end{center}
\caption{Summary of results for the skyrmions called here h-sk and e-sk for both phases considered in this work. Results for
filling factor $\nu=+1$ are obtained from the $\nu=-1$ by exchanging hole and electron due to the
particle-hole symmetry.}
\label{results}
\end{table}

In the case of $\nu=-1$ the hole skyrmion does not feel the anisotropies and is an infinite size spin-polarized valley skyrmion.
The denser electron skyrmion on the contrary has a different behavior in the AF and F phases. In the F phase it is also
fully polarized and infinite-sized so essentially similar to the hole case. In the AF phase the electron skyrmion is still of infinite
size but is unpolarized at small Zeeman energy. A critical value of the ratio of Zeeman energy versus Coulomb energy
is necessary to obtain a fully polarized state. This is in agreement with the measurements of ref.(\onlinecite{Young2012})
at $\nu=-1$ that give evidence for spin flips as seen by a gap increase with increasing total magnetic field. This means
that these measurements are in the AF state.

In the case of $\nu=0$ the counting of states from generalized skyrmion theory is still correct but there is a change
of behavior for the ground state quantum numbers. Indeed we find that for large enough system size the skyrmion is always
in a maximally localized state in real space with a size of the order of a magnetic length. In the F phase the skyrmion is fully spin polarized
while in the AF phase there is a nontrivial magnetization process and a finite Zeeman energy is required to obtain a fully spin-polarized
quasiparticle. Again the polarization mechanism is disconnected from the spatial extent of the skyrmion.
The competition between anisotropies and Zeeman effect means also that there is no longer a universal scaling
of skyrmion physics with the ratio of Zeeman energy to Coulomb energy. Our results are summarized in Table (\ref{results}).

Recent experiments~\cite{Young2012} have measured transport gaps as a function of the total magnetic field applied to the sample
and also as a function of the perpendicular component of the field by using a tilted-field configuration. This allows to separately control
the Coulomb energy (sensitive only to the perpendicular component of the field) and the Zeeman energy (sensitive to the total field).
The transport gap is expected to be due to a skyrmion-antiskyrmion pair excitation. The field dependence of the gap at $\nu=-1$
has been measured~\cite{Young2012} and is steeper as a function of the total field than expected for a simple electron-hole pair. This is
evidence for extra spin flips involved in the excitations. This is exactly what we expect from our results provided the sample is in the AF phase~:
see Table (\ref{results}). The AF phase is also the explanation of the conductance transition observed at very large fields~\cite{Young2014}
so this is a coherent picture. In the anisotropy model we use, it is important to note that contrary to the standard clean skyrmion
model there is no universal scaling as a function of the ratio of Zeeman energy to Coulomb energy. Indeed the Coulomb energy is replaced
by the anisotropy energy (provided it is small enough). Deviations from the standard clean skyrmion scaling have been observed
in ref.(\onlinecite{Young2012}) but only for $\nu=-4$. It remains to be seen if this is also true in the central Landau level.
We have obtained in section IV an estimate for the magnetic field required to fully polarize the skyrmions. If we ask that for the largest
field used in ref.(\onlinecite{Young2012}), $B=35T$, the skyrmion is not fully polarized this requires $g_L\mu_B (B=35T) \lesssim
0.05 (g/\ell_B^2)$ hence an anisotropy energy scale $ g/\ell_B^2 \gtrsim 800K$. Such a value is large but still below
 the Coulomb energy scale $\approx 1500K$ using $\epsilon =2.5$ as appropriate for graphene on hexagonal boron nitride substrate.
The idea behind the introduction of the simple local model Eq.(\ref{aniso}) that anisotropies
are a small perturbation that selects the true ground state within the SU(4) manifold is thus still valid. 
This estimate calls for more attempts to study the magnitude of the anisotropies, notably to find an estimate of the $\theta$ parameter
distinguishing $g_\perp$ and $g_z$.

While the diagonalizations we have performed give information about skyrmion quantum numbers, they cannot lead to reliable estimates
of the associated gaps, but only orders of magnitude. In addition it is hard to access the whole range of  $\theta$ parameter
because there are additional degeneracies at high-symmetry points like the F/AF boundary. A strategy would be to construct
Hartree-Fock wavefunctions that can be evaluated for very large systems and with the correct quantum numbers. We have found
that this is not easy because simple generalizations of Eq.(\ref{n0wavefn}) are not sensitive to anisotropies.
While we have concentrated on the central Landau levels, there is also skyrmion physics in the higher Landau levels that requires
detailed investigation. Another open question is the role of Landau level mixing which is certainly quantitatively important
in monolayer graphene. This is an effect which cannot be straightforwardly included in ED studies since the Fock space is already
very large due to the spin and valley degeneracies so adding more levels is not feasible. However such effects can be treated 
in the Hartree-Fock approximation~: see e.g. ref.(\onlinecite{Roy2014,Feshami2016}). To implement this approach one has to write down
skyrmion wavefunctions generalizing Eq.(\ref{Dsk}). While this will change energetics it will not change our
most important finding i.e. the decoupling of the magnetization process and the spatial profile of the skyrmions.


\begin{acknowledgments}
We acknowledge discussions with Michel Bauer and Inti Sodemann. 
Thanks are due to Takahiro Mizusaki for collaboration at an early stage of this project.
Calculations were performed on the Cobalt computer operated by GENCI-CCRT.
\end{acknowledgments}


\end{document}